\documentclass[12pt]{article}
\let\section=\subsection     \let\subsection=\subsubsection                
\usepackage{graphicx}

\begin{document}
\begin{center}
   {\large \bf HBT INTERFEROMETRY AND THE PARTON-HADRON 
               PHASE TRANSITION}\\[2mm]
   SVEN SOFF \\[5mm]
   {\small \it  Lawrence Berkeley National Laboratory \\
   Nuclear Science Division 70-319, 1 Cyclotron Road, Berkeley, CA94720, USA\\[8mm] }
\end{center}

\begin{abstract}\noindent
   We discuss predictions for the pion and kaon  
   interferometry measurements in relativistic heavy ion collisions 
   at SPS and RHIC energies. In particular, we confront 
   relativistic transport model calculations that include explicitly  
   a first-order phase transition from a thermalized quark-gluon plasma 
   to a hadron gas with recent data from the RHIC experiments. 
   We critically examine the {\it HBT-puzzle} both from the 
   theoretical as well as from the experimental point of view. 
   Alternative scenarios are briefly explained. 
\end{abstract}

\section{Introduction}
This contribution is mainly based on results presented in Refs.\ \cite{soffbassdumi,
kaonlett,sqm2001}. We will briefly summarize the main conclusions 
obtained in these articles and then focus on a critical discussion 
of these results and their comparison to experimental data 
\cite{STARpreprint,Johnson:2001zi} 
and look for possible solutions (of the {\it HBT-puzzle}).

One motivation to study two-particle correlations at small relative momenta 
is due to their predicted sensitivity to a phase transition from quark-gluon matter 
to hadronic matter \cite{pratt86,schlei,dirk1,reviews,Gyulassy:1989yr}. 
In particular, for a first-order phase transition, larger 
hadronization times were expected to lead to considerably enhanced interferometry radii, 
characterizing the space-time extension of the particle-emitting source, compared 
to, for example, a purely hadronic scenario. 
Moreover, one is highly interested in the properties of such a phase transition 
as the critical temperature $T_c$ or the latent heat. 
The radii should also depend on the initial specific entropy density or the initial 
thermalization time of the quark-gluon phase.  

Here, we discuss relativistic transport calculations at RHIC energies that describe the initial 
dense  stage 
by hydrodynamics \cite{DumRi} and the later more 
dilute stages by microscopic transport \cite{bass98,soff00} of the particles.
The two models are matched at the hadronization hypersurface \cite{hu_main}. 
In the hadronic phase the particles are allowed to rescatter and to excite resonances 
based on cross sections as measured 
in vacuum. One example, the K$\pi$ cross section, is shown in Fig.~1. 
For the initial dense (hydrodynamical) phase of a QGP a bag model equation of state 
exhibiting a first-order phase transition is employed. 
Hence, a phase transition in local equilibrium that proceeds through the formation of a mixed phase, 
is considered. The details of this relativistic hybrid transport model 
can be found elsewhere \cite{hu_main}. 
\begin{center}
\vspace*{-1.3 cm}
  \includegraphics[width=10cm,height=8cm]{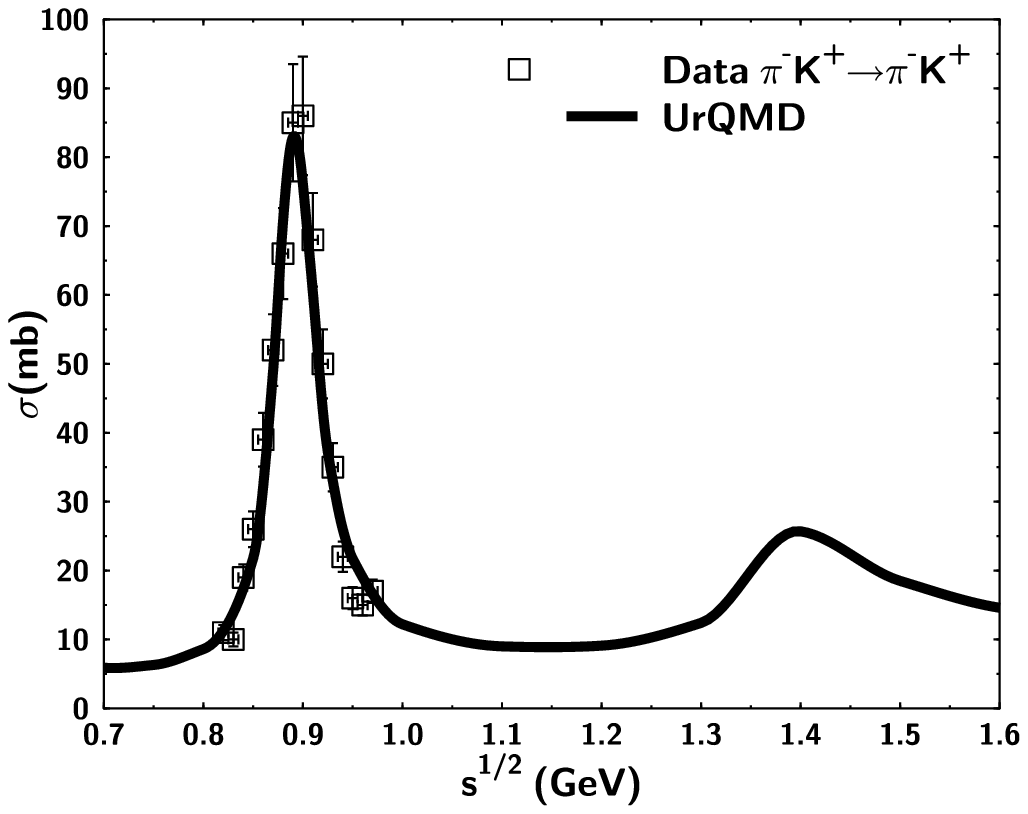}\\
   \parbox{12cm}
        {\footnotesize Fig.~1: Measured and modeled K$^+ \pi^-$ cross section as a 
function of the center-of-mass energy $\sqrt{s}$. The large peak shows the K$^*(892)$ resonance.}
\end{center}
Fig.~2 shows a typical space-time evolution within this model. The contour lines of 
the freeze-out hypersurfaces of pions extend to rather large radii and times compared 
to the  size of the mixed phase. In this hadronic phase many soft collisons take place 
that hardly modify the single-particle spectra but have a strong impact on the 
correlation functions that measure the final freeze-out state. 
\begin{center}
\vspace*{-0.8cm}
  \includegraphics[width=8.5cm,height=6.5cm]{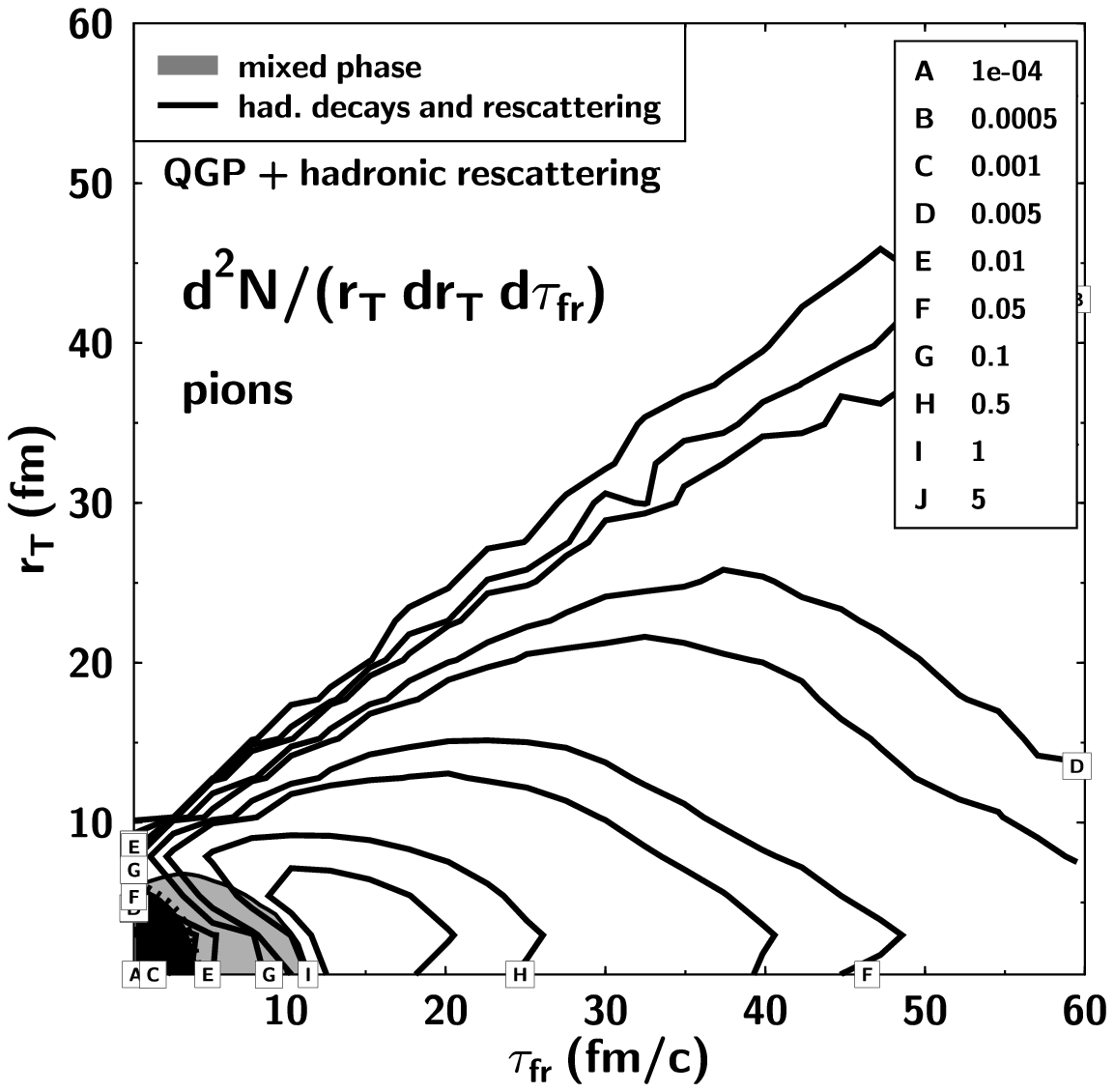}\\
  
   \parbox{12cm}
        {\footnotesize
        Fig.~2: Transverse radius - time plane showing contour lines of the freeze-out
hypersurfaces. The grey-shaded area shows the extension of the mixed phase.}
\end{center}
Studying this in detail lead to the following conclusions \cite{soffbassdumi}: 
(i)  The dissipative hadronic phase leads to a rather large duration of emission.
(ii) The $R_{\rm out}/R_{\rm side}$ ratio, thought to be a characteristic measure of this 
emission duration, increases with transverse momentum.
(iii) The specific dependencies of the interferometry radii on the QGP properties are rather weak 
due to the dominance of the hadronic phase. This even leads to qualitative 
differences if calculations with and without this subsequent hadronic phase are compared 
(dependence on the critical temperature).
\section{Why kaons?}
The kaon correlations provide a severe test of the pion data and have several 
advantages \cite{kaonlett,sqm2001,Gyulassy:1989yr}. 
In particular, the kaon density is much lower than the pion 
density \cite{Murray:2002ek}. 
Hence, multiparticle correlations that might play a role for the pions are 
of minor importance for the kaons. Also, the contributions from long-lived resonances are under 
better control for kaons. 
The $R_{\rm out}/R_{\rm side}$ ratio for kaons is shown in Fig.~3. 
Most important is the strongly increased sensitivity to $T_c$ and the specific entropy density 
(SPS vs.\ RHIC) at larger transverse momenta ($K_T \sim 1\,$GeV/c). 
This enhanced sensitivity is also driven by a strong increase of the direct emission component 
(from the phase boundary) at high $K_T$ as shown in Fig.~4. 
More and more kaons (up to $\sim 30\%$) escape  
the initial stages (unperturbed by the hadronic phase). 
\begin{center}
  \includegraphics[height=7cm]{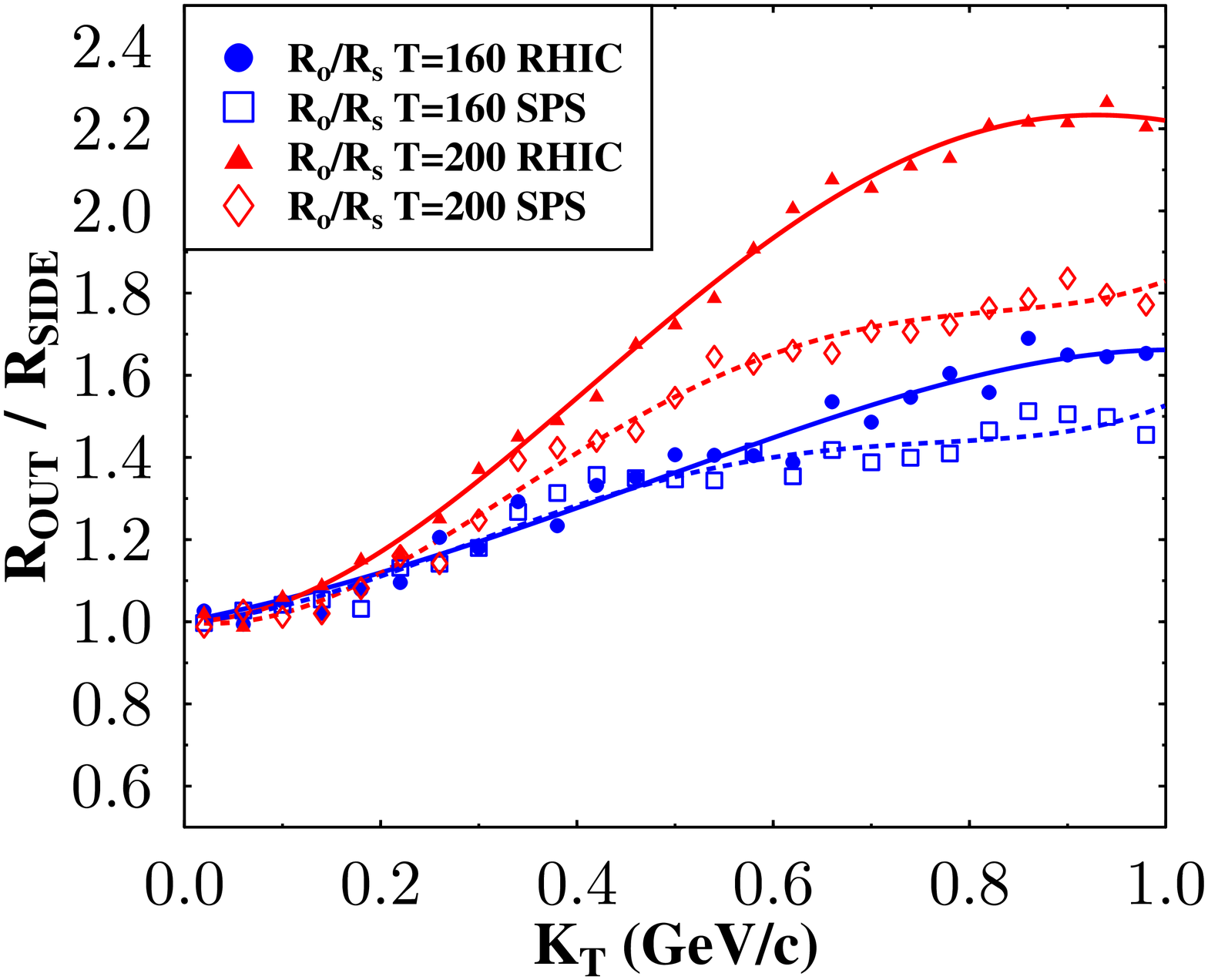}\\
   \parbox{12cm}
        {\footnotesize
        Fig.~3: $R_{\rm out}/R_{\rm side}$ for kaons at RHIC (full symbols) and
at SPS (open symbols), as a function of $K_T$ for critical temperatures
$T_c\simeq 160\,$MeV  and $T_c\simeq 200\,$MeV, respectively.}
\end{center}

\begin{center}
  \includegraphics[width=10cm,height=8cm]{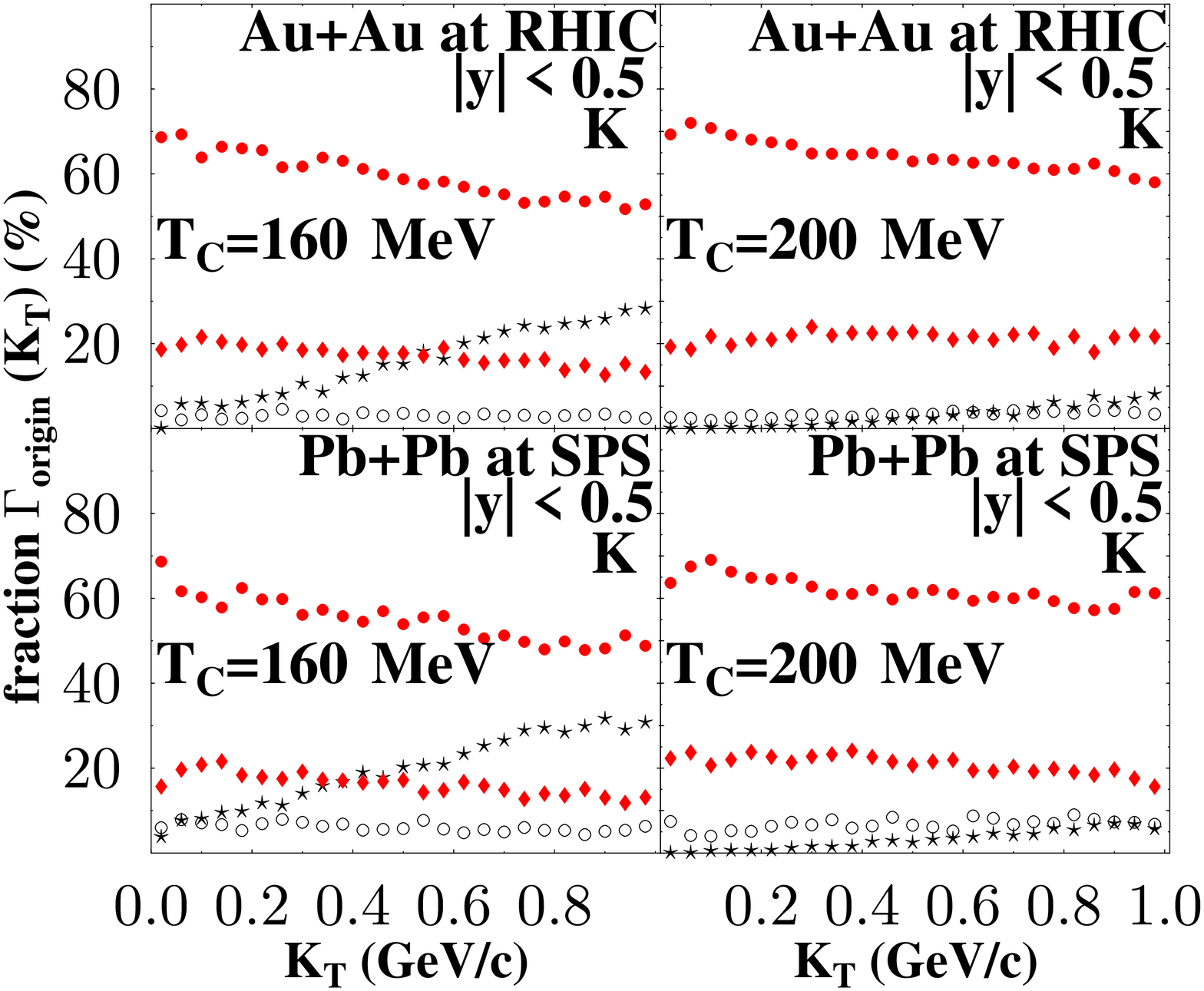}\\ 
   \parbox{12cm}
        {\footnotesize
        Fig.~4: Fraction of kaons $\Gamma_{\rm origin}$ that origin from a particular
reaction channel prior to freeze-out. These are resonance
decays (full circles),  direct emission from the phase boundary (stars),
elastic meson-meson (diamonds), or elastic meson-baryon (open circles)  
collisions. The upper and lower diagrams are for RHIC 
and SPS initial conditions for
$T_c\simeq 160\,$MeV (left) and $T_c\simeq 200\,$MeV (right), respectively.
}
\end{center}
Fig.~5 shows the correlation parameters $R_{\rm o},\,R_{\rm s},\,R_{\rm l}\,$and $\, \lambda$ 
as obtained from the explicit calculation of the correlation functions 
\cite{Pratt:1990zq} in the 
respective 
transverse momentum bins and subsequent fitting of these correlation functions to 
a Gaussian form of the correlator $C_2=1+\lambda \exp(-R_{\rm o}^2q_{\rm o}^2
-R_{\rm s}^2q_{\rm s}^2-R_{\rm l}^2q_{\rm l}^2)$.  
Most important, we learn that even for a first-order phase transition scenario 
the interferometry radii are not unusually large. The transverse radii $R_{\rm o}$ and $R_{\rm s}$ 
are less than $7\,$fm. Moreover, we also note a strong effect of a finite momentum 
resolution (fmr) that has to be corrected for in the experimental analysis. 
The radii and the $\lambda$ intercept parameter are reduced by the fmr. The reduction is 
stronger for higher $K_T$. Experimental data for the kaons 
from the RHIC experiments will be available soon.  
At the moment, the pion data are of great interest.  
\section{Pion interferometry radii - theory versus data}
The table shows the experimental STAR data (average of $\pi^-$ and $\pi^+$ data plus
generous error bars) \cite{STARpreprint} and the results of fitting the 3-dimensional
correlation functions (as obtained from the transport calculations 
(RHIC initial conditions, $T_c\approx 160\,$MeV) + {\it correlation
after burner} (by Pratt) \cite{Pratt:1990zq}) to a Gaussian correlator.
\begin{center}
  \includegraphics[width=10cm,height=8cm]{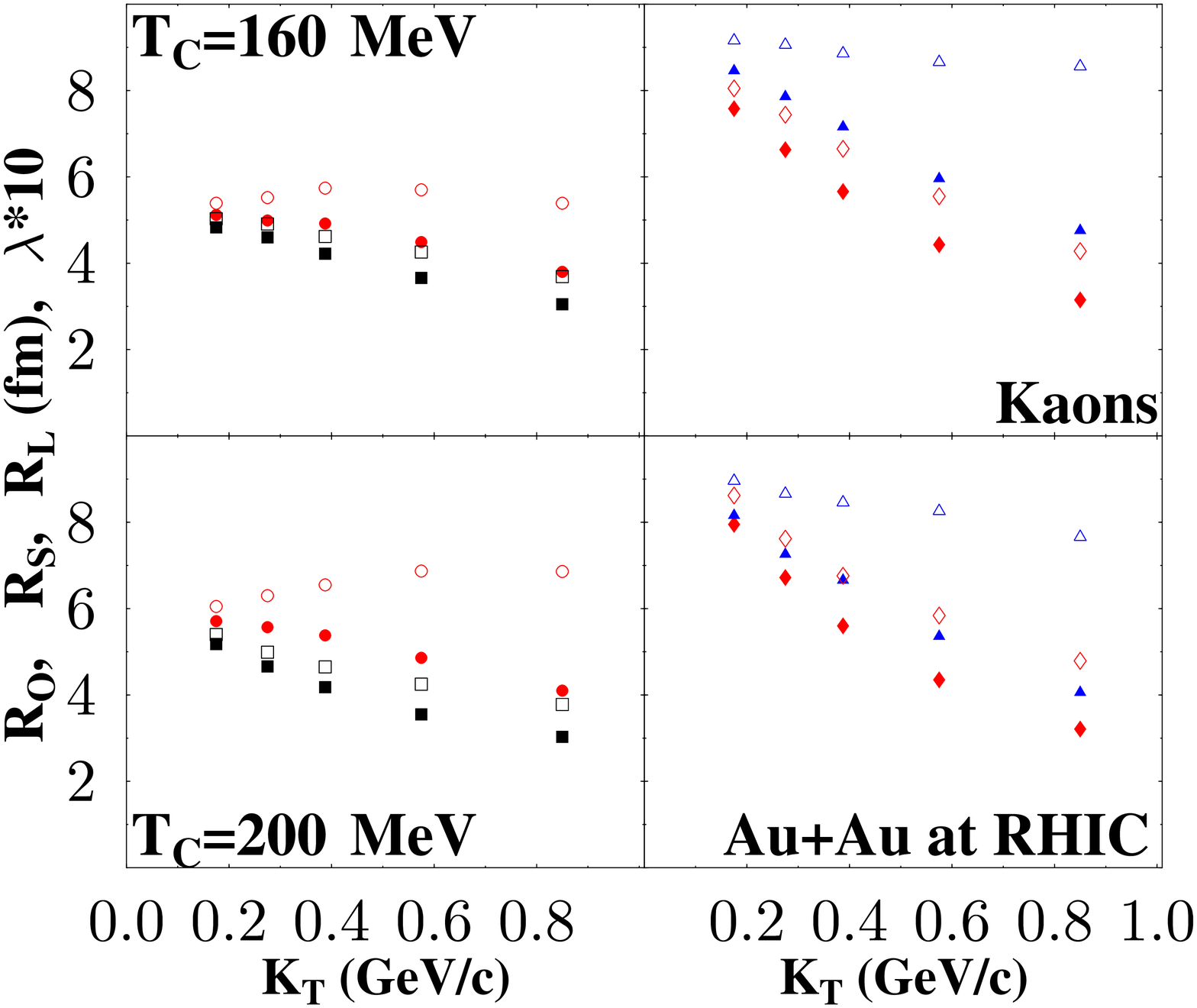}\\ 
   \parbox{12cm}
        {\footnotesize
        Fig.~5: Kaon HBT-parameters $R_{\rm out}$ (circles),
$R_{\rm side}$ (squares),
$R_{\rm long}$ (diamonds) and $\lambda \cdot 10$ (triangles) 
as obtained from a $\chi^2$ fit of $C_2$  to an  
Gaussian {\it ansatz}  
for Au+Au collisions at RHIC as calculated with $T_c \simeq 160\,$MeV (top) and
$T_c \simeq 200\,$MeV (bottom). Full and open circles correspond to calculations
with and without taking momentum resolution effects into account,
respectively.}
\end{center}
\begin{center}
\begin{tabular}{|l|r||l|l|r|}
\hline
     &      & STAR & T  & T$_{\rm fmr}$\\
\hline
low & $R_{\rm out}$  & $ 5.9 \pm 0.5 $  &  $8.2   $    & $7.4$ \\
$K_T 1$ & $R_{\rm side}$ & $5.7 \pm 0.5$ &  $5.7   $    & $5.4$ \\
 &$R_{\rm out}/R_{\rm side}$ & $1.04 $  & $ 1.4  $    & $1.4$\\
\hline
med. &$R_{\rm out}$  & $ 5.3 \pm 0.6 $  &  $8.0   $    & $6.8$ \\                     
$K_T 2$ &$R_{\rm side}$ & $5.3 \pm 0.5$ &  $5.3   $    & $4.8$ \\                        
 &$R_{\rm out}/R_{\rm side}$ & $ 1.0 $  & $ 1.5  $    & $1.4$\\
\hline
high & $R_{\rm out}$  & $ 4.5 \pm 0.6 $  &  $7.7   $    & $6.1$ \\                     
$K_T 3$ &$R_{\rm side}$ & $5.1 \pm 0.6$ &  $4.8  $    & $4.3$ \\                        
 &$R_{\rm out}/R_{\rm side}$ & $ 0.88$  & $ 1.6  $    & $1.4$\\
\hline
\end{tabular}
\end{center}
Since the data are corrected they should be compared to the calculations (T) 
that do not take into account fmr. 
While the $R_{\rm side}$ radii appear to be described even too good the $R_{\rm out}$ radii 
are too large compared to the data.
However, these pion radii are considerably smaller than the corresponding radii obtained 
from the coordinate-space points and using expressions for the Gaussian radius parameters based 
on a saddle-point integration over 
the source function \cite{soffbassdumi,Hardtke:2000vf}.  
Only the values presented in the table which are obtained 
from the complete calculation 
and the performed fits should be compared to data. 
The $R_{\rm out}/R_{\rm side}$ ratio is also larger 
than unity for the fitted values and 
confirms the so-called {\it HBT-puzzle}, i.e., 
the RHIC data from STAR and PHENIX \cite{STARpreprint,Johnson:2001zi} 
indicate a decreasing ratio with $K_T$ (even below 1) 
and all calculations show only ratios larger 1.  
Note however, that the exp.\ data are consistent in 
the sense that they can be described by a 
single set of fit parameters \cite{snellings}. 
On the other hand, there is presently no dynamical 
transport model describing this trend. 
\section{Discussion of the {\it HBT-puzzle}}
\subsection{Experimental Uncertainties and SPS data}
The following list provides an overview which corrections enter the experimental data analysis. 
All of them are thought to be under relatively good control and accounted for 
in the systematic error bars. 
The numerous corrections illustrate the difficult, complex and challenging task to extract 
the {\it true} correlation parameters from the {\it raw} data.
Without further commenting we list the corrections \cite{STARpreprint}  
(i) two-track resolution, 
(ii) particle identification (electrons, contributions from 
weak decays, e.g. $\Lambda$),
(iii) track splitting (one particle interpreted as two), 
(iv) track merging (two particles interpreted as one, requirement of seperated tracks, 
affects low $q$ pairs and reduces radii), 
(v) Coulomb corrections (should be under good control, except maybe influence of weak 
decay pions), 
(vi) momentum resolution (strong $K_T$ dependent correction, reduces radii, see, e.g.,  Fig.~5), 
(vii) collider mode peculiarities (collision vertex, i.e., acceptance region varies event by event)

At the CERN-SPS, pion \cite{Bearden:1998aq,wa98,na49,appels,wa97} 
and kaon \cite{na44prlnew,na49new} 
interferometry have been investigated for 
Pb+Pb collisions. These data (see, e.g., \cite{wa98,na49}) seem to support 
an increasing $R_{\rm out}/R_{\rm side}$ ratio 
with $K_T$, being larger than 1. 
This would mean a real qualitative change of the reaction 
dynamics from SPS to RHIC energies.
\subsection{Model assumptions and uncertainties}
Next, we provide a list of theoretical assumptions and uncertainties,  
that is by far not complete.  
(a) 
the  hadronization process itself is usually modelled via a prescription, 
(b) the binary collision approximation (here only used for the later dilute stages after hadronization) 
requires in principle sufficiently low particle densities, 
(c) the approximation {\it ideal} fluid dynamics (local thermal distributions, no dissipation) 
is certainly questionable, in particular if applied for the later dilute stages, 
(d) the limitations of {\it pure} hydrodynamical calculations as for example given due to the 
(pre/de)scription of the freeze-out,  
(e) the choice of the hadronization hypersurface to {\it switch} between models needs 
to be further elaborated (although it seems to be somehow a natural choice due to the 
limitations of the individual models),  
(f) the assumption of cylindrical symetric transverse expansion and  longitudinal 
scaling flow (should be justified at midrapidity and high energies), 
(g) the role of in-medium effects (both on the hadron properties or the equation of state), 
(h) the assumption that nucleation proceeds via hadronic bubbles ({\it well-mixed} phase scenario) 
may be questioned [see below],
(i) the large number of hadronic states in the model equation of state may 
not be realized (they speed up the hadronization ($\tau_H\sim \tau_is_i/s_H(T_c)$)) 
and reduce the time-delay signal. So neglecting them should even increase the 
observed differences. 

Fig.~6, taken from Ref.\ \cite{Zschiesche:2001dx} by Zschiesche et al., 
addresses several uncertainties. First of all, the dependence on the 
latent heat (strong first-order vs.\ weak first-order vs.\ cross-over phase transition 
with a vanishing latent heat) is examined. Secondly, the effect of varying the 
choice of the freeze-out temperature $T_f$ is investigated. 
\begin{center}
\vspace*{-1.0cm}
  \includegraphics[width=14cm,height=8cm]{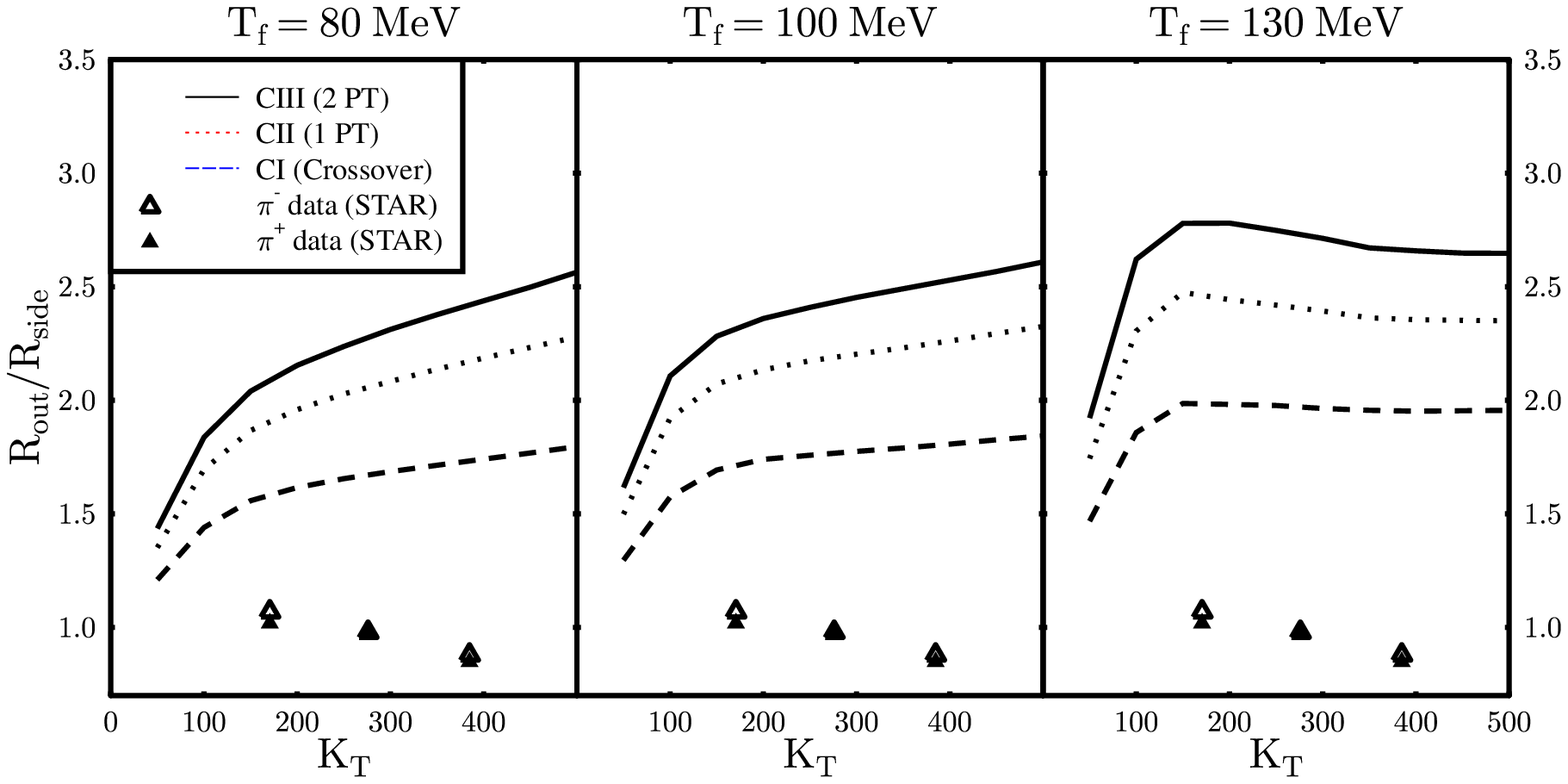}\\
   \parbox{12cm}
        {\footnotesize
        Fig.~6: Figure taken from \cite{Zschiesche:2001dx} (Zschiesche et al.).
{\it Pure} hydrodynamical calculations
with a strong first-order (solid line), a weaker first-order (dotted line), and
a cross-over (dashed line) phase transition and different freeze-out 
temperatures $T_f$.}
\end{center}
Although reducing the latent heat reduces the large $R_{\rm out}/R_{\rm side}$ ratios, 
the  $K_T$ dependence of the data cannot be described. Lowering the freeze-out temperature 
also does not  help in this approach. 
Even rather exotic initial conditions, for example, a collective flow
prior to a typical equilibration time of the order of $\tau_i\sim 1\,$fm/c,
do not provide a notable improvement \cite{Heinz:2001xi}.
What could help is a strongly opaque source (see also \cite{sqm2001,hwbarz}) that 
suppresses the spatial component in the {\it out} radius compared to the {\it side} radius 
(leaving out of the discussion the role of $xt$-correlations). 

A completely different model scenario (illustrated in Fig.~7) is existent if  
the large expansion rate of the system leads to strong supercooling below a 
{\it spinodal} temperature $T_S$ such that 
the system disintegrates rather instantaneously \cite{spino}. 
In this case of a {\it spinodal instability} 
the soft mixed phase vanishes. As a result the reaction times are quite short what should be 
reflected in the interferometry radii and their ratio. This has to be estimated 
in a quantitative way. 
\begin{center}
  \includegraphics[width=8cm,height=8cm]{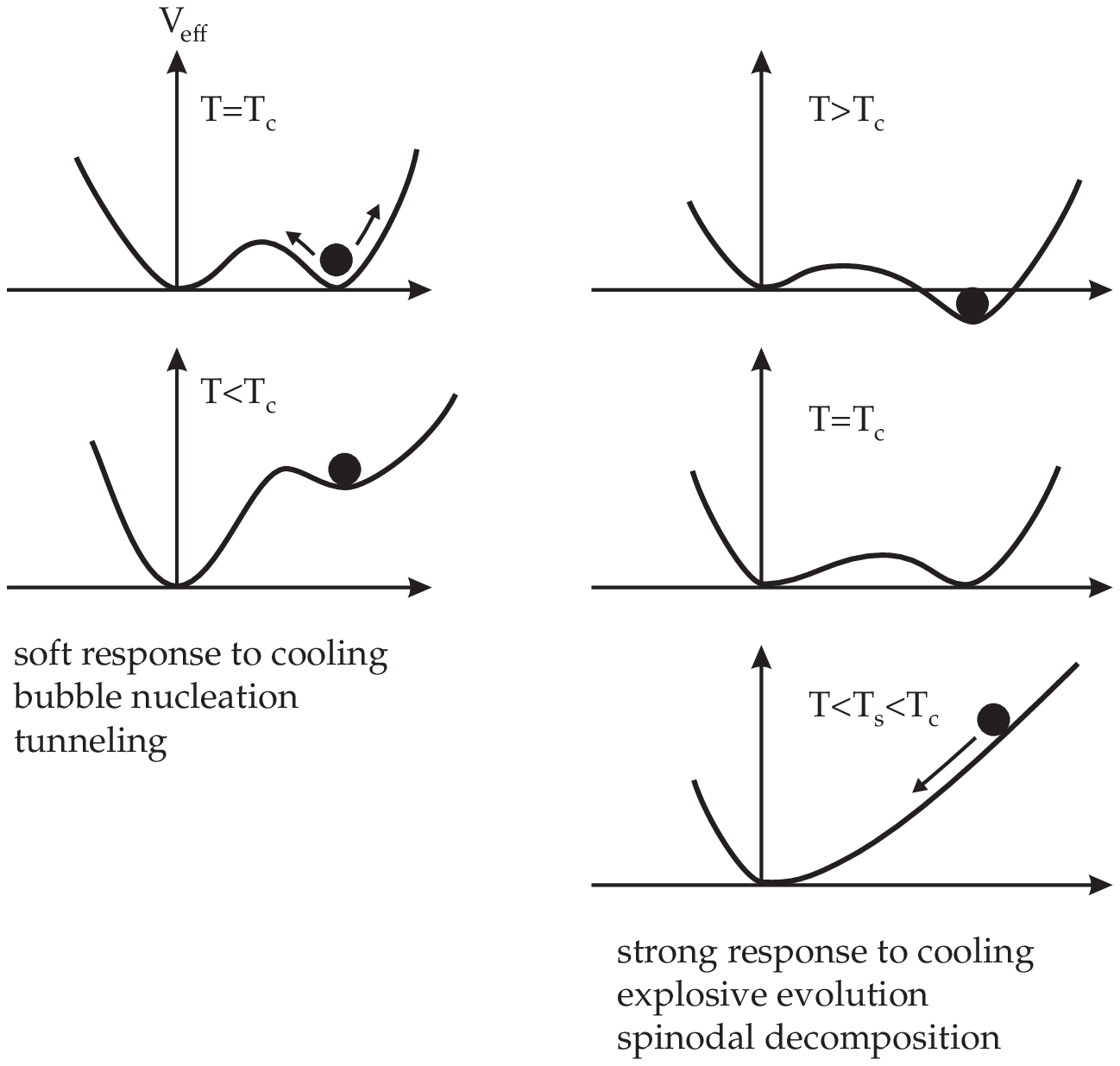}\\
  
   \parbox{12cm}
        {\footnotesize
        Fig.~7: Illustration of the effective potential for temperatures
        around the critical temperature $T_c$ in the case of a
        {\it slow} nucleation via a mixed phase ({\it bubble nucleation})
        (left)
        and in an {\it explosive} scenario that leads to supercooling down to
        a spinodal temperature $T_s$ at or below which the system disintegrates rather
        instantaneously (right).
}
\end{center}
\section{Summary}
We demonstrated that the kaon interferometry measurements, in particular 
at high $K_T$,  
will provide an exellent probe of the space-time dynamics 
(close to the phase boundary). 
In addition they represent a severe test of the pion correlations 
and may help to better understand the {\it HBT-puzzle}, i.e., the 
difference in the $K_T$ dependence of the $R_{\rm out}/R_{\rm side}$ ratio 
between the model predictions and the experimental (RHIC) data. 
A closer look showed us that the differences are due to the  
$R_{\rm out}$ radii  (which are larger in the model calculations) 
while the $R_{\rm side}$ radii seem to be described reasonably. 
Finally, we discussed some possible origins of these differences.

\vspace*{0.5cm}

\noindent{\bf Acknowledgments}

\noindent
SS would like to thank the organizers of this very active and productive workshop for invitation.
SS is supported by the Alexander von Humboldt Foundation through a 
Feodor Lynen Fellowship and DOE Grant No.\ DE-AC03-76SF00098.
SS is grateful to S.\ Bass, A.\ Dumitru, D.\ Hardtke, and S.\ Panitkin 
for fruitful collaborations.

\end{document}